# Observation of spin-orbit magnetoresistance in metallic thin films on magnetic insulators

Lifan Zhou,[1,†] Hongkang Song,[2,3,†] Kai Liu,[3] Zhongzhi Luan,[1] Peng Wang,[1] Lei Sun,[1] Shengwei Jiang,[1] Hongjun Xiang,[3,4] Yanbin Chen,[1,4] Jun Du,[1,4] Haifeng Ding,[1,4] Ke Xia,[2] Jiang Xiao,[3,4,5,*] and Di Wu[1,4,*]

[1] National Laboratory of Solid State Microstructures and Department of Physics, Nanjing University, Nanjing 210093, P. R. China.
[2] Department of Physics, Beijing Normal University, Beijing 100875, P. R. China.
[3] Department of Physics and State Key Laboratory of Surface Physics, Fudan University, Shanghai 200433, P. R. China.
[4] Collaborative Innovation Center of Advanced Microstructures, Nanjing 210093, P. R. China.
[5] Institute for Nanoelectronics Devices and Quantum Computing, Fudan University, Shanghai 200433, P. R. China.
† These authors contributed equally to this work.
*Corresponding author: xiaojiang@fudan.edu.cn, dwu@nju.edu.cn.

**Abstract**

A magnetoresistance effect induced by the Rashba spin-orbit interaction was predicted, but not yet observed, in bilayers consisting of normal metal and ferromagnetic insulator. Here, we present an experimental observation of this new type of spin-orbit magnetoresistance (SOMR) effect in a bilayer structure Cu[Pt]/$Y_3Fe_5O_{12}$ (YIG), where the Cu/YIG interface is decorated with nanosize Pt islands. This new MR is apparently not caused by the bulk spin-orbit interaction because of the negligible spin-orbit interaction in Cu and the discontinuity of the Pt islands. This SOMR disappears when the Pt islands are absent or located away from the Cu/YIG interface, therefore we can unambiguously ascribe it to the Rashba spin-orbit interaction at the interface enhanced by the Pt decoration. The numerical Boltzmann simulations are consistent with the experimental SOMR results in the angular dependence of magnetic field and the Cu thickness dependence. Our finding demonstrates the realization of the spin manipulation by interface engineering.

**Introduction**

Relativistic spin-orbit interaction (SOI) plays a critical role in a variety of interesting phenomena, including the spin Hall effect (SHE) (*1-3*), topological insulators (*4*), the formation of skyrmions (*5, 6*). In SHE, a pure spin current transverse to an electric current can be generated in conductors with strong SOI, such as Pt, Ta etc (*7, 8*). The inverse SHE (ISHE) is generally used to detect the spin current electrically by converting a pure spin current into a charge current (*9, 10*). It was recently discovered that the interplay of the SHE and ISHE in a nonmagnetic heavy metal (NM) with strong SOI in contact with a ferromagnetic insulator (FI) leads to an unconventional magnetoresistance (MR) - the spin Hall magnetoresistance (SMR), in which the resistance of the NM layer depends on the direction of the FI magnetization *M* (*11-13*). SMR has been observed in several NM/FI systems and even in metallic bilayers (*14-17*). However, it has been argued that SMR may originate from the magnetic moment in the NM layer induced by the magnetic proximity effect (MPE) (*18*). These two mechanisms were proposed to be distinguished by the angular dependent MR measurements (*11, 13*). Very recently, another type of MR, the Hanle MR (HMR), is demonstrated in a single metallic film with strong SOI owing to the combined actions of SHE and Hanle effect (*19*). HMR depends on the direction and the strength of the external magnetic field $H$, rather than that of $M$ in SMR. Within the framework of SMR, because of the negligible SOI in Cu (*20*), one would not expect any MR effect in a Cu/FI bilayer.

Recently, Grigoryan *et al*. predicted a new type of MR effect in the NM/FI systems when a Rashba type SOI is present at the interface between NM and FI (*21*). This new spin-orbit MR (SOMR) works even with light metals such as Cu or Al with negligible bulk SOI, provided that the Rashba SOI is present at the NM/FI interface. Because of the identical angular dependence on $M$ direction for SOMR and SMR, however, it is difficult to distinguish SOMR from SMR in systems like Pt/$Y_3Fe_5O_{12}$ (YIG), where both SOMR and SMR are present in principle. In this work, we report the first observation of SOMR in a Cu/YIG bilayer, where the Rashba SOI at Cu/YIG interface is enhanced by an ultrathin Pt layer (< 1 nm). We also confirmed that SOMR almost disappears when Pt is placed inside or on the other side of the Cu layer, indicating that SMR from the ultrathin Pt layer cannot be the origin of the observed MR and the Pt-decoration of the Cu/YIG interface is crucial for SOMR. The observed SOMR has the same angular dependence as the SMR in Pt/YIG, in agreement with the SOMR prediction (*21*). The monotonous Cu-thickness dependence of SOMR is clearly different from the non-monotonous dependence of SMR (*13*, *18*). Both the angular- and Cu-thickness- dependence of the observed MR are in good agreement with our Boltzmann simulations based on the SOMR mechanism. In addition, the MR shows two maxima as the Pt layer thickness increases, in sharp contrast with that of SMR (*13*, *22*).

## Results and discussions
### Sample morphology and structure
The YIG films used in this study are 10 nm thick, unless otherwise stated, grown by pulsed laser deposition (PLD) on $Gd_3Ga_5O_{12}$ (GGG) (111) substrates. The surface morphology of the YIG films was characterized by atomic force microscopy (AFM), as shown in Fig. 1A. The film is fairly smooth with the root-mean-square (rms) roughness of 0.127 nm and the peak-to-valley fluctuation of 0.776 nm. The 0.4-nm-thick Pt layer, thinner than the peak-to-valley value of the YIG film, deposited on YIG by magnetron sputtering forms the nanosize islands with the rms roughness of ~ 0.733 nm, shown in Fig. 1B. This discontinuous Pt layer is non-conductive with the resistance over the upper limit of a multimeter. The surface roughness is reduced after the deposition of Cu onto Pt, as shown in Fig. S1. Figure 1C presents the cross-section high-resolution transmission electron microscope (HRTEM) image of the Au(3)/Cu(4)[Pt(0.4)]/YIG films, where the numbers are the thicknesses in the unit of nanometer. The YIG film is clearly single-crystalline and smooth. The lattice constant of the YIG film is determined to be 1.2234 nm, to be compared to 1.2366 nm for the bulk YIG. A clear interface is observed between the metallic films and the YIG film. The metallic films are polycrystalline.

### Field-dependent magnetization and transport measurements
In this work, all the measurements were performed at room temperature. The YIG film is almost isotropic in the film plane with the coercivity of about 0.4 Oe, shown in Fig. 2A. Due to the large paramagnetic background of the GGG substrate, it is difficult to measure the magnetization of a thin YIG/GGG film in the out-of-plane geometry. We measured a 400-nm-thick YIG/GGG(111) film instead. As shown in Fig. 2B, the magnetization is saturated at ~ 1800 Oe. The saturation magnetization $M_s$ of our YIG film is determined to be 164.5 emu/cc measured by ferromagnetic resonance (FMR) (see the Supplementary Materials). In comparison, $M_s$ of bulk YIG is 140 emu/cc.

Figures 2C and 2D present the resistivity $\rho$ as a function of $H$ for Cu(2)[Pt(0.4)]/YIG(10) sample. In experiments, $H$ was applied along i) the direction of the current $I$ ($x$-axis), ii) in the sample plane and perpendicular to the current direction (*y*-axis), and iii) perpendicular to the sample plane (*z*-axis), respectively. The MR effects are clearly present in all measurements. For *H* along *x*- and *y*-directions, $\rho$ shows two peaks around the coercive fields of YIG. For *H* along *z*-direction, $\rho$ shows a minimum at $H = 0$ and remains almost a constant value above the saturation field. These features indicate that the MR effects are intimately correlated with *M*, meaning that the observed MR effects are not HMR.

**Angular dependent MR measurements**
To further study the anisotropy of the MR effects in Cu[Pt]/YIG, we performed the angular dependent MR measurements. Figure 3A shows $\Delta\rho/\rho$ of Cu(3)[Pt(0.4)]/YIG(10) sample with rotation of *H* in the *xy*- ($\alpha$-scan), *yz*- ($\beta$-scan) and *xz*- ($\gamma$-scan) planes, where $\alpha$, $\beta$ and $\gamma$ are the angles between *H* and *x*-, *z*- and *z*-directions, respectively, as defined in the inset of Fig. 3A. The applied magnetic field strength ($H = 1.5$ T) is large enough to align *M* with *H*. The MR effect is clearly anisotropic. The MR ratio, defined as $\Delta\rho/\rho = [\rho(\text{angle}) - \rho(\text{angle} = 90^\circ)]/\rho(\text{angle} = 90^\circ)$, in $\alpha$- and $\beta$-scans is about 0.012%, comparable to the SMR ratio in Pt/YIG (see Fig. S3) (*11*, *13*, *23*).

Next, we investigated the origin of the observed MR effect. Considering that Pt on YIG may suffer from the MPE induced ferromagnetic moment and the corresponding anisotropic MR (AMR) (*24*), we replaced Pt by a 0.4-nm-thick Au layer, which is well-known to have a negligible MPE (*25*). The MR effect of 0.002% still appears as shown in Fig. 3B, comparable to the SMR ratio in Au/YIG (see Fig. S4), ruling out MPE as the origin of the observed MR. Furthermore, the MR ratios of the Cu(3)[Pt(0.4)]/YIG(10) sample in $\alpha$- and $\beta$-scans are comparable and almost one order of magnitude larger than that in $\gamma$-scan. This is different from AMR of a ferromagnetic metal, where the MR ratio in $\alpha$- and $\gamma$-scans is much larger than that in $\beta$-scan (*11*, *14*, *24*). Therefore, the MPE-induced AMR can be ruled out.

In fact, the behaviors of the MR angular dependence follow the SMR scenario well (*11*, *13-15*, *17*). However, with several control experiments, we can unambiguously exclude SMR as the explanation for our observations.

First, the observed MR amplitude cannot be explained by SMR. In our samples, the 0.4-nm-thick ultrathin Pt layer is non-conductive and the conductivity of bulk Pt is about one order of magnitude smaller than that of bulk Cu, meaning that the current mainly passes through the Cu layer. We prepared a 3-nm-thick single layer Cu on YIG without interface decoration and performed the MR angular dependent measurement in $\alpha$-scan. MR is not observed, as shown in Fig. 3B, evidencing that the Pt-decorated interface is indispensable. A conductive 0.4-nm-thick Pt layer is not available experimentally. Considering that a small fraction of current may flow in the Pt islands, there is a possibility of the occurrence of SMR from the Pt islands. According to the reported SMR results in Pt/YIG bilayers, the SMR ratio in Pt/YIG decreases rapidly with decreasing Pt thickness when the Pt thickness is less than about 3 nm (*13*, *18*). The SMR ratio of Pt(0.4)/YIG is extrapolated to be well below 0.01% from the previously reported $\Delta\rho/\rho$ versus Pt thickness data (*13*, *18*). Considering the pronounced shunting current of the highly conductive Cu layer, the SMR ratio should be significantly reduced in Cu[Pt]/YIG, *i.e.*, much less than 0.01%. In comparison, the MR ratio is as large as ~ 0.012% in Cu(3)[Pt(0.4)]/YIG (see Fig. 3A). Therefore, the SMR mechanism cannot explain our observations.

Second, the potential enhancement of SMR caused by intermixing or alloying between a strong SOI material and a weak SOI material can be excluded (*17*, *26*, *27*). For this purpose, we prepared two types of control samples with the 0.4-nm-thick Pt layer either on top of or inserted inside the Cu layer: [Pt(0.4)]Cu(3)/YIG and Cu(1)[Pt(0.4)]Cu(3)/YIG. Since both samples are fabricated under the same condition as the Cu[Pt]/YIG samples, the intermixing of Pt and Cu should be similar. The MR vanishes in the [Pt(0.4)]Cu(3)/YIG and Cu(1)[Pt(0.4)]Cu(3)/YIG samples, shown in Fig. 3B. These results rule out the Pt-Cu alloying induced SMR. Thus, we conclude that the observed MR effect is not SMR.

**Cu-thickness dependent transport measurements**
To identify the physical origin of the observed unusual MR, we carried out the Cu-thickness dependent measurements. Figure 4A presents the angular dependent MR measurements of Cu($t_{Cu}$)[Pt(0.4)]/YIG in $\alpha$-scans for various Cu thickness ($t_{Cu}$). Obviously, the MR ratio steadily decreases with increasing $t_{Cu}$, highlighting the importance of the Pt-decorated Cu/YIG interface. This monotonous NM-thickness dependence of this MR is in sharp difference with the non-monotonous behavior of SMR, which peaks at ~ 3 nm for Pt/YIG (*13*, *18*). The Cu-thickness dependence of $\rho$ and the MR ratio extracted from Fig. 4A are shown in Fig. 4B. For very thin Cu film ($t_{Cu}$ ≤ 5 nm), $\rho$ dramatically increases with decreasing $t_{Cu}$, indicating that $\rho$ is dominated by the interface/surface scatterings.

Besides SMR, there is another type MR predicted recently possessing the same angular dependence as we found (see Fig. 3A) (*21*). It originates from the Rashba SOI at the interface of a NM/FI bilayer. By comparing the samples of Cu[Pt]/YIG, [Pt]Cu/YIG and Cu[Pt]Cu/YIG, one can see that only the Cu[Pt]/YIG samples exhibit a significant MR (see Fig. 3B). It strongly suggests that the MR observed in our experiments is the SOMR predicted in Ref. *21*, and the Pt-decoration enhances the Rashba SOI at the Cu/YIG interface.

**First principles calculations and Boltzmann simulations**
In order to prove that the Pt-decoration can indeed induce Rashba SOI at the Cu/YIG interface, we carried out first principles band structure calculations based on i) a Cu ultra-thin film of 14 monolayers, ii) the same Cu film as i) but covered by Au on surfaces on both sides, iii) the same Cu film as i) but covered by Pt on both surfaces, iv) the Pt layer inside the Cu film. By comparing these four different scenarios, we can see that there is no clear Rashba effect in the bare Cu film and the one covered by Au. A strong Rashba effect appears only for Pt on the Cu film surface (the details of calculations are given in the Supplementary Materials).

For a quantitative analysis, we employ a Boltzmann formalism to calculate the charge and spin transport in a NM/FI bilayer structure. We solve the following spin-dependent Boltzmann equation in the NM layer:

$$\mathbf{v}_0(\mathbf{k}) \cdot \frac{\partial f_\alpha(\mathbf{r},\mathbf{k})}{\partial \mathbf{r}} - e\mathbf{E} \cdot \mathbf{v}_0(\mathbf{k}) \, \delta_{\alpha,0} = -R_\alpha(\mathbf{k}) f_\alpha(\mathbf{r},\mathbf{k}) + \sum_{\alpha'=0,x,y,z} \int_{FS} d\mathbf{k}' P_{\alpha,\alpha'}(\mathbf{k},\mathbf{k}') f_{\alpha'}(\mathbf{r},\mathbf{k}'), \quad (1)$$

where $f_{\alpha=0,x,y,z}(\mathbf{r},\mathbf{k})$ is the four-component distribution function denoting the charge/spin occupation at position **r** and wavevector **k**. The interface at FI $z = z_+$ contains a Rashba type SOI described by the Hamiltonian: $H_R = \eta \hat{\boldsymbol{\sigma}} \cdot (\hat{\mathbf{z}} \times \hat{\mathbf{p}}) \delta(z - z_+)$, where $\eta$ is the strength of the Rashba SOI, $\hat{\mathbf{z}}$ is the normal direction of the interface, $\hat{\mathbf{p}}$ is the momentum operator. $H_R$ gives rise to an anomalous velocity localized at the interface. The Boltzmann equation is solved by discretizing the spherical Fermi surface of Cu and the real space in *z* direction of Cu film. With the full distribution function, we calculate all charge/spin transport properties, including the longitudinal and transverse conductivities. This method extends the earlier Boltzmann method developed for current-perpendicular-to-plane structure like spin valves to current-in-plane structure like NM/FI bilayers (*28-31*) by taking into account the surface roughness and Rashba SOI at the interface. The details of the simulations are given in the Supplementary Materials.

In the numerical Boltzmann calculation, there are only two fitting parameters, the surface roughness and the Rashba coupling constant. All other parameters are either given by the experiment (such as the film thickness) or can be determined otherwise (such the bulk relaxation time). By employing the quantum description of rough surface (*32-34*), we are able to fit the thickness dependence of $\rho$ in the ultra-thin Cu film to a reasonably good precision as shown in Fig. 4B. It is quite surprising considering that there is only one fitting parameter – the surface roughness. Once surface roughness is determined, we calculate the magnetization angular dependence of $\rho$, in a good agreement with the experimental results (see Fig. 3A), from which we obtain the SOMR ratio. The calculated SOMR ratio is shown in Fig. 4B, which shows monotonic decreasing behavior as a function of Cu film thickness, consistent with our experiment results but very different from the non-monotonic behavior observed in SMR (*13, 18*).

**Pt-thickness dependent MR measurements**
Finally, to further differentiate SOMR from SMR, we carried out the Pt thickness $t_{Pt}$ dependent measurements. To reduce the sample fluctuation, we fabricated the YIG films successively under the same condition. Figure 4A shows the angular dependent MR measurements of Cu(3)/Pt($t_{Pt}$)/YIG in $\alpha$-scans with $H$ = 2000 Oe. The MR ratio extracted from Fig. 4A exhibits non-monotonous behavior with increasing $t_{Pt}$ as shown in Fig. 4C. Two separate regimes can be identified: 1) the SOMR regime for $t_{Pt}$ < 1 nm and 2) the conventional SMR regime for $t_{Pt}$ > 2.2 nm (see Fig. 4C). For $t_{Pt}$ < ~0.6 nm, $\rho$ and $\Delta\rho/\rho$ increase with increasing $t_{Pt}$ because the Pt islands not only introduce the interface scattering but also enhance the Rashba SOI. For ~0.6 nm < $t_{Pt}$ < ~1 nm, the Pt islands start to form a complete layer, leading to the reduction of the interface roughness and the rapid decrease of $\rho$ as seen in Fig. 4C. The MR ratio continues to increase in this region because of the enhanced Rashba SOI with increasing Pt coverage on YIG. For ~1 nm < $t_{Pt}$ < ~2 nm, $\rho$ is smaller than the resistivity of Cu/YIG, suggesting that the interface scattering has minor contribution to $\rho$. Since SOMR is caused by the interface scattering, the MR ratio rapidly drops in this region. A sizable SMR ratio only appears when $t_{Pt}$ > 2 nm in Pt/YIG (*13, 18, 22*). Therefore, around $t_{Pt}$ ~ 2 nm, both SOMR and SMR are small, resulting in a minimum in MR. In the SMR regime, the SMR ratio exhibits a maximum, as expected for SMR (*13, 18, 22*). This result demonstrates the differences between the SOMR and the SMR.

A theoretical calculation shows that a rough interface can enhance SHE (*34*). To understand the role of the roughness to the SOMR, we fabricated a control sample of Cu(3)[Ag(0.7)]/YIG. The rms roughness of Ag(0.7)/YIG is 0.797 nm, as shown in Fig. S9(A), similar as that of Pt(0.4 nm)/YIG. Owing to the weak SOI in Ag, the Rashba SOI in Cu[Ag]/YIG is expected to be weak. We do not observe any MR effect down to $5\times10^{-6}$ in Cu(3)[Ag(0.7)]/YIG, shown in Fig. S9(B). This result means that the rough surface alone cannot cause the SOMR.

**Conclusions**
In conclusion, we report the first observation of the SOMR effect predicted recently (*21*) at room temperature in Cu/YIG films with the Pt decoration at interface. We show that this MR effect is caused by the enhanced Rashba SOI at the Pt-decorated interface. The angular dependence of SOMR is similar to that of SMR, but all other features are different, such as the increasing MR with decreasing Cu thickness. The amplitude of the SOMR ratio is comparable to that of the SMR ratio in Pt/YIG, highlighting the importance of the NM/FI interfaces. Our finding demonstrates the possibility of realizing spin manipulation by interface decoration.

**Materials and Methods**
    The single crystalline YIG films were epitaxially grown on GGG (111) substrates by PLD technique using a KrF excimer laser with wavelength of 248 nm. The PLD system was operated at a laser repetition rate of 4 Hz and an energy density of 10 J/cm$^2$. The distance between the substrate and the target is 50 mm. Before films deposition, the chamber was evacuated to a base pressure of $1 \times 10^{-7}$ torr. The YIG films were deposited at ~ 730 $^o$C in an oxygen pressure of 0.05 Torr. The growth of the YIG films was monitored by in situ reflection high-energy electron diffraction (RHEED). The structure was further examined by X-ray diffraction and HRTEM. The magnetic properties of all YIG films were characterized using a vibration sample magnetometer (VSM). Then we used magnetron sputtering to deposit polycrystalline metallic films onto the YIG films via dc sputtering at room temperature with a shadow mask to define 0.3-mm-wide and 3-mm-long Hall bars. The deposition rate was calibrated by X-ray reflectivity. After the metallic film deposition, the samples were immediately mounted and transferred into a vacuum chamber for the transport measurements to minimize the metal oxidation. The resistance was measured by a Keithley 2002 multimeter in a four-probe mode. For the angular dependent MR measurements with the magnetic field less than 5000 Oe, the resistance was monitored as the magnet was rotated. The angular dependent MR measurements with the magnetic field larger than 5000 Oe were performed in a physical property measurement system (PPMS) equipped with a rotatory sample holder.

**H2: Supplementary Materials**
    section S1. AFM images of Cu($t_{Cu}$)[Pt (0.4)]/YIG(10)/GGG(111)
    section S2. Magnetic properties of the YIG films
    section S3. SMR in Pt/YIG
    section S4. SMR and AFM image of Au/YIG
    section S5. First principles calculations
    section S6. Boltzmann simulations
    fig. S1. AFM images of Cu($t_{Cu}$)[Pt(0.4)]/YIG(10)/GGG (111).
    fig. S2. FMR of the YIG films.
    fig. S3. SMR in Pt/YIG.
    fig. S4. SMR and AFM image of Au/YIG.
    fig. S5. The band structures of Cu, Au/Cu/Au and Pt/Cu/Pt.
    fig. S6. The spin textures of outer band and inner band.
    fig. S7. The Rashba splitting in Cu/Pt/Cu.
    fig. S8. Specular and diffusive interface scattering in the NM/FI bilayer.
    fig. S9. AFM image of Ag(0.7)/YIG and MR of Cu(3)[Ag(0.7)]/YIG.
    References (*35–52*)

**Acknowledgments: Funding:** L.Z. and D.W. are supported by National Key R&D Program of China (2017YFA0303202), NSF of China (11674159, 51471086 and 11727808), National Basic Research Program of China (2013CB922103). H.S. and J.X. acknowledge the support by NSF of China (11474065 and 11722430) and National Key R&D Program of China (2016YFA0300702). **Author contributions:** J.X. and D.W. designed and supervised the project. L.F.Z. and Z.Z.L. prepared the samples. L.F.Z. performed the transport measurements with support from Z.Z.L., P.W., S.W.J. H.K.S. performed the Boltzmann simulations under supervision of J.X. K.L. performed the first principles calculations under supervision of H.J.X. L.S. and Y.B.C. were responsible for the HRTEM characterization. J.X., D.W. and L.F.Z. wrote the manuscript and J.D., H.J.X., H.F.D. and K.X. commented on the manuscript. All authors discussed the results and reviewed the manuscript. **Competing interests:** The authors declare no competing financial interests. **Data and materials availability:** All data needed to evaluate the conclusions in the paper are present in the paper and/or the Supplementary Materials. Additional data related to this paper may be requested from the authors.

**Figures and Tables**

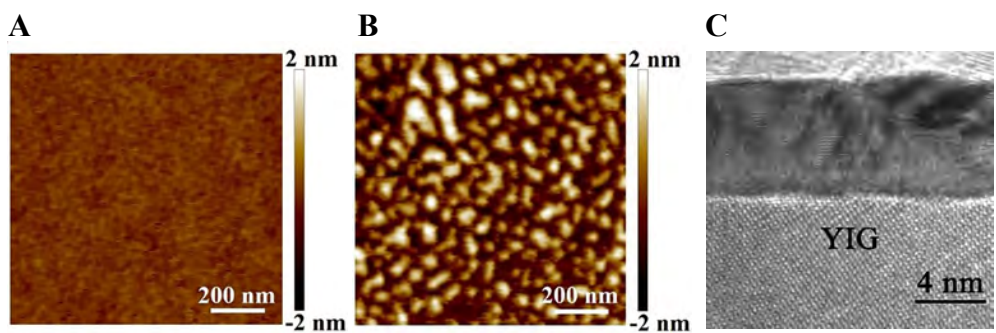

**Fig. 1. Sample characterization. (A)** AFM image of YIG(10)/GGG, the rms roughness is 0.127 nm. **(B)** AFM image of Pt(0.4)/YIG(10)/GGG, the rms roughness is 0.733 nm. **(C)** HRTEM image of Au(3)/Cu(4)[Pt(0.4)]/YIG heterostructure where Au is used to prevent the oxidation.

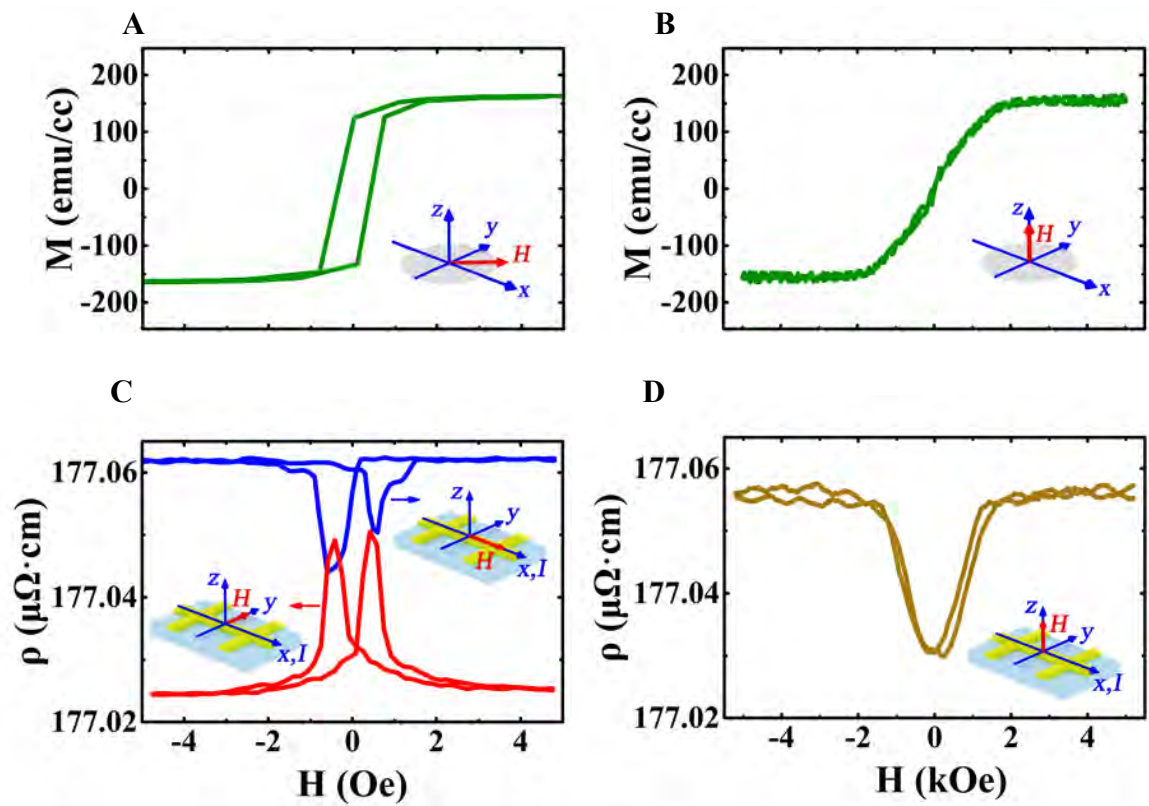

**Fig. 2. Field-dependent magnetization and transport measurements.** Magnetic hysteresis loops of **(A)** YIG(10)/GGG with field in-plane and **(B)** YIG(400)/GGG with field out-of-plane. $\rho$ measured on the Cu(2)[Pt(0.4)]/YIG(10)/GGG sample for $H$ applied **(C)** along $x$-axis, $y$-axis and **(D)** $z$-axis, respectively.

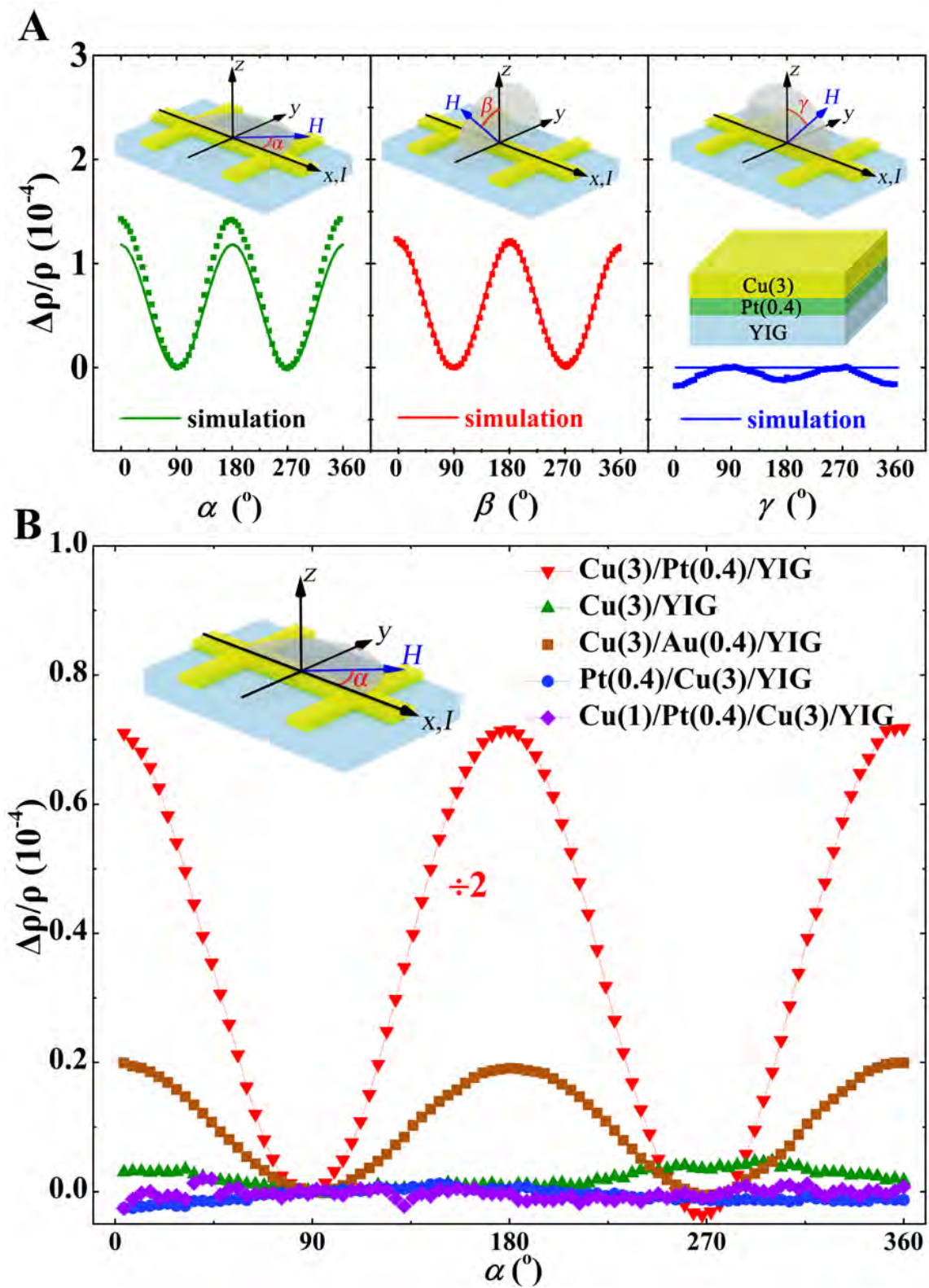

**Fig. 3. Angular dependent MR measurements.** (**A**) Angular dependent MR measurements in the *xy*, *yz*, and *xz* planes for Cu(3)[Pt(0.4)]/YIG. The solid lines are the Boltzmann simulation results. (**B**) Angular dependent MR measurements in the *xy* plane for several control samples.

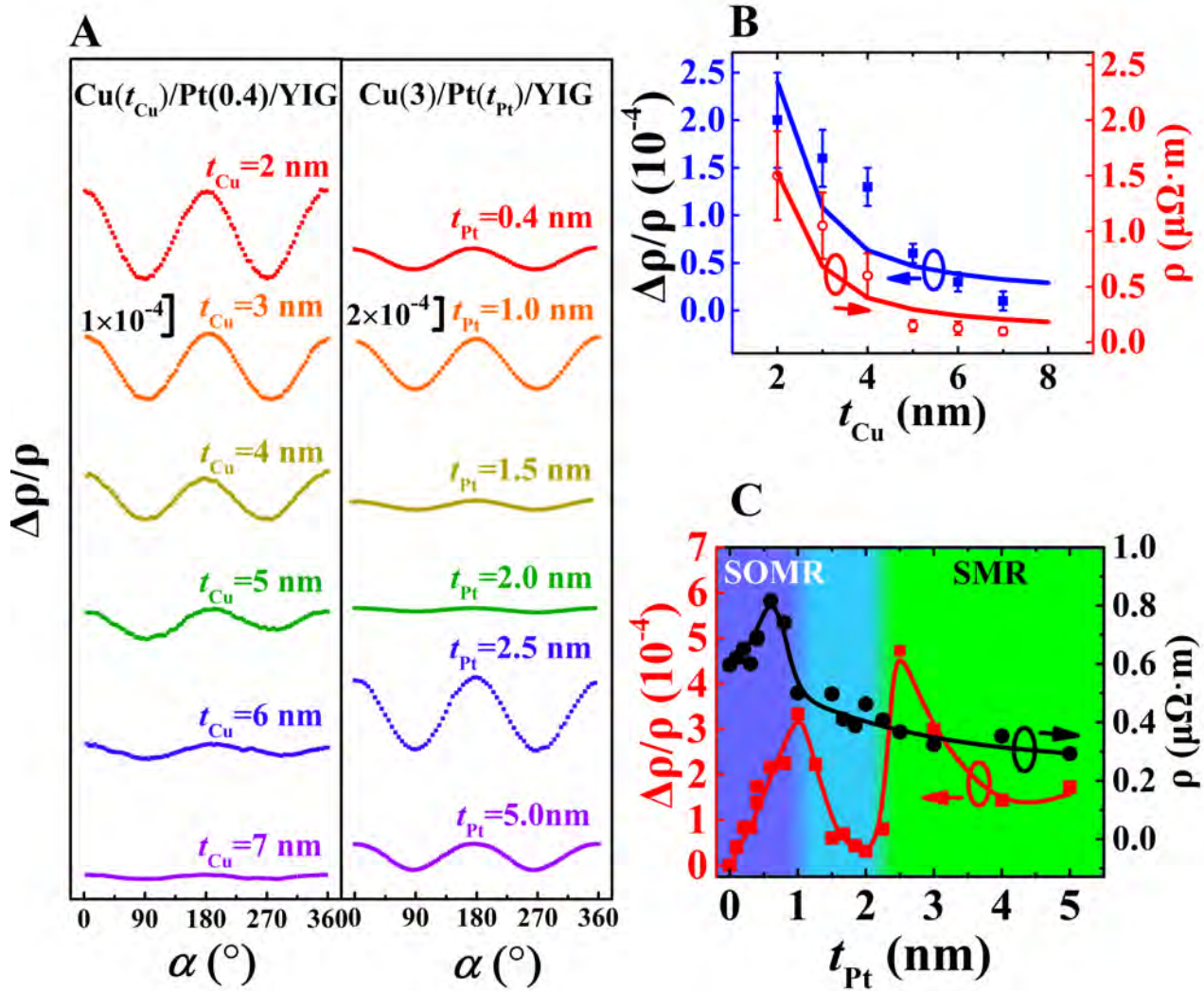

**Fig. 4. Cu- and Pt-thickness dependent transport measurements.** Angular dependent MR measurements in the $xy$ plane for **(A)** Cu($t_{Cu}$)[Pt(0.4)]/YIG samples and Cu(3)/Pt($t_{Pt}$)/YIG samples. **(B)** Cu thickness dependence of the MR ratio and $\rho$, respectively, for Cu($t_{Cu}$)[Pt(0.4)]/YIG. The solid lines are the Boltzmann simulation results. **(C)** The Pt layer thickness dependence of the MR ratio and $\rho$, respectively, for Cu(3)/Pt($t_{Pt}$)/YIG. The solid lines are guide to the eyes.

# Supplementary Materials for

## Observation of spin-orbit magnetoresistance in metallic thin films on magnetic insulators

Lifan Zhou, Hongkang Song, Kai Liu, Zhongzhi Luan, Peng Wang, Lei Sun, Shengwei Jiang, Hongjun Xiang, Yanbin Chen, Jun Du, Haifeng Ding, Ke Xia, Jiang Xiao, and Di Wu

**This PDF file includes:**

- section S1. AFM images of Cu($t_{Cu}$)[Pt (0.4)]/YIG(10)/GGG(111)
- section S2. Magnetic properties of the YIG films
- section S3. Spin Hall magnetoresistance in Pt/YIG
- section S4. SMR and AFM image of Au/YIG
- section S5. First principles calculations
- section S6. Boltzmann simulations
- fig. S1. AFM images of Cu($t_{Cu}$)[Pt(0.4)]/YIG(10)/GGG (111).
- fig. S2. FMR of the YIG films.
- fig. S3. Spin Hall magnetoresistance in Pt/YIG.
- fig. S4. SMR and AFM image of Au/YIG.
- fig. S5. The band structures of Cu, Au/Cu/Au and Pt/Cu/Pt.
- fig. S6. The spin textures of outer band and inner band.
- fig. S7. The Rashba splitting in Cu/Pt/Cu.
- fig. S8. Specular and diffusive interface scattering in the NM/FI bilayer.
- fig. S9. AFM image of Ag(0.7)/YIG and MR of Cu(3)[Ag(0.7)]/YIG.
- References (*35–52*)

## section S1. AFM images of Cu($t_{Cu}$)[Pt (0.4)]/YIG(10)/GGG(111)

The discontinuous 0.4-nm-thick Pt layer is insulating. The surface morphology gets smoother after the deposition of Cu onto Pt. The rms roughness of Cu($t_{Cu}$)[Pt(0.4)]/YIG(10)/GGG(111) increases with increasing $t_{Cu}$, shown in Fig. S1. For the thin Cu film, resistivity $\rho$ increases with decreasing Cu thickness $t_{Cu}$, indicating that $\rho$ is dominated by the interface/surface scatterings.

## section S2. Magnetic properties of the YIG films

The saturation magnetization $M_s$ is measured by ferromagnetic resonance (FMR) with an in-plane magnetic field and in an X-band microwave cavity operated at a frequency of $f =$ 9.7798 GHz. Fig. S2A shows the FMR absorption derivative spectrum of the YIG (10 nm) film measured at room temperature. The resonance field $H_r$ is at 2608.8 Oe. $M_s$ is determined to be 164.5 emu/cc by using the Kittel formula: $f = \gamma\sqrt{H_r(H_r + 4\pi M_S)}$ where $\gamma$ is the gyromagnetic ratio. In comparison, $M_s$ of bulk YIG is 140 emu/cc.

## section S3. Spin Hall magnetoresistance in Pt/YIG

Pt/YIG is a typical system with the SMR. We fabricated a sample of Pt(3.4)/YIG to compare with the magnetoresistance of Cu[Pt]/YIG. Fig. S3 presents the angular dependent magnetoresistance of our Pt(3.4)/YIG sample measured in $\beta$-scan at room temperature. The applied magnetic field $H = 1.5$ T is much larger than the demagnetization field to align the magnetization along $H$. We determined the SMR ratio of about $4.5\times10^{-4}$, comparable to previous reports (*11*, *14*).

## section S4. SMR and AFM image of Au/YIG

It would be better to compare the MR ratio of the Cu[Au]/YIG sample with Au/YIG sample. Fig. S4A shows the angular dependent magnetoresistance of the Au(6)/YIG sample measured in α-scan at room temperature. A SMR ratio of ~$1.2\times10^{-5}$ is observed, consistent with the previous reports (*17*). The SMR ratio in Au/YIG should be larger for the optimized Au thickness. The surface of the Au film is smoother than that of the Pt film, shown in Fig. S4B. Therefore, the observed magnetoresistance ratio of about $2\times10^{-5}$ in Cu(3)[Au(0.4)]/YIG is reasonable.

## section S5. First principles calculations

The calculations are performed within density-functional theory (DFT) using the projector augmented wave (PAW) method (*35*) encoded in the Vienna ab initio simulation package (VASP) (*36*, *37*). The exchange-correlation potential is treated in the generalized-gradient approximation (GGA) (*38*). The plane-wave cutoff energy is set to be 400 eV. For geometry optimization, all the internal coordinates are relaxed until the Hellmann-Feynman forces are less than 1meV/Å and SOI is not included. For the band structure calculation, the SOI is included.

We build three models. The first model is a pure Cu ultra-thin film of 14 monolayers. The second (third) one is the same film covered by Au (Pt) at surfaces on both sides to keep the inversion symmetry of the whole system. The thickness of the bare Cu film, the vacuum layer and the surface lattice constant are 27 Å, 20 Å and 2.56 Å, respectively.

The band structures of the three models are shown in Fig. S5. There is no obvious Rashba splitting in the bare Cu film and in the film covered by Au, as shown in Fig. S5A and S5B, respectively. While in the Cu film covered by Pt, there is a Rashba splitting, and the splitted bands are highlighted by green bold lines, as shown in Fig. S5C. Near the Gamma point, the bands highlighted by green bold lines are very similar to the parabolic energy dispersion of a two-dimensional-gas in a structure inversion asymmetric environment, characteristics of the *k*-linear Rashba effect. To further confirm the nature of the Rashba splitting in Fig. S5C, we calculate the spin textures of outer band and inner band around -0.35 eV and -0.10 eV iso-energy surface, respectively, shown in Fig. S6. The inverse rotation of spin orientations of outer band and inner band is characteristic of a pure Rashba splitting. This indicates that Pt can indeed induce a strong Rashba effect at Cu surfaces.

In Fig. S7, we show that when Pt is placed inside Cu film away from the surface, the Rashba splitting decreases significantly, and vanishes when Pt is in the middle of the film. This is consistent with our experimental data that when Pt is placed inside Cu, the SOMR disappear.

## section S6. Boltzmann simulations

Based on the Boltzmann method developed for CPP (current-perpendicular-to-plane) structure like spin valves (*29-31, 39, 40*), we made modifications for the CIP (current-in-plane) structure like the bilayer systems used in SMR/HMR/SOMR.

### 6.1 Basic formalism of Boltzmann calculation

We use four-component distribution function $f_{\alpha=0,x,y,z}(\mathbf{r},\mathbf{k})$ to denote the charge/spin occupation at position $\mathbf{r}$ and wavevector $\mathbf{k}$: $f_0$ is the electric charge distribution and $\boldsymbol{f} = (f_x, f_y, f_z)$ is the pure spin (no net charge) distribution. Thus, the majority/minority spin distribution $f_\pm = f_0 \pm |\boldsymbol{f}|$, where $\pm$ denotes the majority/minority spin along $\pm \hat{\boldsymbol{f}}$ direction. These four-component distribution function satisfies the generalized spin-dependent Boltzmann equations (*31, 39-41*),

$$\mathbf{v}_0(\mathbf{k}) \cdot \frac{\partial f_\alpha(\mathbf{r},\mathbf{k})}{\partial \mathbf{r}} - e\mathbf{E} \cdot \mathbf{v}_0(\mathbf{k}) \, \delta_{\alpha,0} = -R_\alpha(\mathbf{k}) f_\alpha(\mathbf{r},\mathbf{k}) + \sum_{\alpha'=0,x,y,z} \int_{FS} d\mathbf{k}' P_{\alpha,\alpha'}(\mathbf{k},\mathbf{k}') f_{\alpha'}(\mathbf{r},\mathbf{k}'), \quad (S1)$$

where $\mathbf{E}$ is the applied external electric field and $\mathbf{v}_0 = \hbar \mathbf{k}/m$ is the velocity in free electron model, and the right-hand side is the scattering-out and scattering-in collision terms and $\int_{FS} d\mathbf{k}'$ denotes the integral over Fermi surface. $R_\alpha(\mathbf{k}) = \sum_{\alpha'} \int_{FS} d\mathbf{k}' P_{\alpha',\alpha}(\mathbf{k}',\mathbf{k})$ is the total relaxation rate for $f_\alpha(\mathbf{r},\mathbf{k})$. And $P_{\alpha,\alpha'}(\mathbf{k},\mathbf{k}')$ describes the $\mathbf{k}' \to \mathbf{k}$ scattering probability from charge/spin-$\alpha'$ to the charge/spin-$\alpha$, e.g., $P_{0,0}(\mathbf{k},\mathbf{k}')$ is the scattering probability for electric charge, $P_{x,x}(\mathbf{k},\mathbf{k}')$ is the spin-conserved scattering probability for spin-$x$, $P_{y,x}(\mathbf{k},\mathbf{k}')$ is the scattering probability with spin flip spin-$x \to$ spin-$y$, $P_{0,x}(\mathbf{k},\mathbf{k}')$ is the scattering probability with spin Hall effect converting pure spin-$x$ to charge, and $P_{x,0}(\mathbf{k},\mathbf{k}')$ is the scattering probability with ISHE converting charge to pure spin-$x$. In the case of normal metal like Cu without bulk SHE, $P_{\alpha,\alpha'}(\mathbf{k},\mathbf{k}') = A_{FS}^{-1}\delta_{\alpha,\alpha'}/\tau(\mathbf{k}')$ in the relaxation time approximation, where $A_{FS}$ is the area of Fermi surface and $\tau(\mathbf{k}')$ is the spin-conserved relaxation time for electrons at $\mathbf{k}'$. $R_\alpha(\mathbf{k})=1/\tau(\mathbf{k})+(1-\delta_{\alpha,0})/\tau_{sf}(\mathbf{k})$, where $\tau_{sf}(\mathbf{k})$ is the spin-flip relaxation time. All scatterings are assumed to be elastic, i.e., $|\mathbf{k}|=|\mathbf{k}'|$.

We study an NM/FI bilayer structure, as shown in Fig. S8, whose interfaces/surfaces are in $x$-$y$ plane and locate at $z_\pm = \pm d/2$. The boundary condition for the upper interface at $z=z_+$ is given by a surface scattering matrix $S^+$ that connects the impinging distribution function ($k_z > 0$) and the reflected distribution function ($k_{z'} > 0$):

$$f_\alpha(z_+,\mathbf{k},k_z > 0) = \int_{FS} d\mathbf{k}' S^+_{\alpha,\alpha'}(\mathbf{k},\mathbf{k}') f_{\alpha'}(z_+,\mathbf{k}',k_z' < 0). \tag{S2}$$

We regard the interface/surface scattering as specular when

$$S^+_{\alpha,\alpha'}(\mathbf{k},\mathbf{k}') = \delta_{\alpha,\alpha'}\delta(k_x - k_x')\delta(k_y - k_y')\delta(k_z + k_z')\Theta(k_z)\Theta(-k_z'), \tag{S3}$$

and as diffusive when

$$S^+_{\alpha,\alpha'}(\mathbf{k},\mathbf{k}') = \delta_{\alpha,\alpha'}A_{FS}^{-1}\delta(|\mathbf{k}|-|\mathbf{k}'|)\Theta(k_z)\Theta(-k_z'), \tag{S4}$$

where the factor $\delta_{\alpha,\alpha'}$ means that the surface scattering is spin-conserving. Similar boundary condition can be written down at $z=z_-$. Due to conservation of charge, we have the following identity

$$\int_{FS} d\mathbf{k} S_{0,0}^{\pm}(\mathbf{k},\mathbf{k}') = \int_{FS} d\mathbf{k}' S_{0,0}^{\pm}(\mathbf{k},\mathbf{k}') = 1 \ . \tag{S5}$$

Since spin is generally not conserved, there is no constraint on the spin related boundary scattering matrix.

Once the distribution function has been found by solving Eq. (S1), all transport properties can be calculated accordingly:

$$\text{charge/spin accumulation:} \ \mu_\alpha(\mathbf{r}) = -e \int_{FS} \frac{d^3\mathbf{k}}{(2\pi)^3} f_\alpha(\mathbf{r},\mathbf{k}), \tag{S6a}$$

$$\text{charge current density } (\alpha=0): j_0^\beta(\mathbf{r}) = -e \int_{FS} \frac{d^3\mathbf{k}}{(2\pi)^3} \left[ v_0^\beta(\mathbf{k}) f_0(\mathbf{r},\mathbf{k}) + \sum_{\alpha=x,y,z} v_\alpha^\beta f_\alpha(\mathbf{r},\mathbf{k}) \right], \tag{S6b}$$

$$\text{spin current density } (\alpha=x,y,z): j_\alpha^\beta(\mathbf{r}) = -e \int_{FS} \frac{d^3\mathbf{k}}{(2\pi)^3} \left[ v_0^\beta(\mathbf{k}) f_\alpha(\mathbf{r},\mathbf{k}) + v_\alpha^\beta f_0(\mathbf{r},\mathbf{k}) \right], \tag{S6c}$$

where $\mu_\alpha$ is the charge accumulation when $\alpha = 0$ and spin-$\alpha$ accumulation when $\alpha = x, y, z$, $j_0^\beta$ is the charge current flowing in $\beta$-direction, $j_\alpha^\beta$ and is the spin-$\alpha$ current flowing in $\beta$ direction when $\alpha = x, y, z$. The two contributions in $j_0$ and $j_\alpha$ are due to the fact that different spins may have different velocities, e.g., the majority/minority spin-$\alpha$ has velocity $\mathbf{v}_0 \pm \mathbf{v}'_\alpha$, where $\mathbf{v}'_\alpha$ is the anomalous velocity due to the spin-orbit coupling [see Eq. (S10) below]. In the bilayer structure in Fig. S7 with translational invariance in x-y plane, $j_\alpha^\beta(\mathbf{r}) = j_\alpha^\beta(z)$, and the film conductivities can be calculated from the current density as

$$\sigma_\alpha^\beta = \frac{1}{E} \int_{z_-}^{z_+} dz j_\alpha^\beta(z). \tag{S7}$$

For electric field applied in x direction, the longitudinal conductivity is $\sigma_0^x$, and the transverse Hall conductivity is $\sigma_0^y$.

To carry out the Boltzmann calculation numerically, we discretize the Fermi surface in **k**-space and the real space (in the out-of-plane z-direction only) simultaneously:

$\{\mathbf{k}_i\}_{i=1}^{n_k}, \{z_j\}_{j=1}^{n_z}$. The Boltzmann equation then becomes a set of linear equations, which is solved by matrix inversion.

### 6.2 Surface roughness

For metallic thin films, the rough surface becomes an important or even dominate factor on the transport. There are various models in dealing with a rough surface, including the Fuchs-Sondheimer model (*42, 43*), the Mayadas-Shatzkes model (*44*), and the Namba model (*45*). All these models are phenomenological, and work only in certain circumstances (*46-51*).

To deal with ultrathin films, we adopt the quantum description of a rough surface as developed in Ref. *32-34*, in which the relaxation rate becomes channel dependent:

$$\frac{1}{\tau_n} \equiv \frac{1}{\tau(\mathbf{k}_{m,n})} = \frac{1}{\tau_0} + \frac{n^2}{\tau'} = \frac{1}{\tau_0} + \frac{1}{\tau'_n} \quad \text{with} \quad \frac{1}{\tau'} = \frac{\delta^2}{a^2} \frac{4S}{3n_c^2} \frac{E_F}{\hbar}, \quad \text{(S8)}$$

where $\tau_0$ is the bulk impurity relaxation time and $\tau'_n$ is the channel-dependent surface relaxation time, and $S = 3\sum_{n'=1}^{n_c} n'^2 / n_c^3 \approx 1$. Here *a* is the lattice constant and $\delta$ parameterizes the magnitude of the surface roughness. In Eq. (S8), the relaxation due to the surface roughness is built into the Boltzmann equation via total relaxation rate as the bulk impurity scattering, rather than a simple surface scattering. The reason for this is the following: for ultrathin film, one cannot view the electron as a classical point particle bouncing back and forth between the two surfaces, and only feels the surface as the electron hits the surface. Instead, the electronic wave function spreads out in the thickness direction and is in contact with the surface all the time, thus the rough surface becomes a 'bulk' effect and is felt constantly by the electron and causes scattering from $\mathbf{k}'$ to $\mathbf{k}$.

### 6.3 Interfacial Rashba spin-orbit interaction

When considering the interfacial Rashba spin-orbit interaction at the NM/FI interface, we assume a Rashba Hamiltonian of the following form (*21*)

$$H_R = \eta \hat{\boldsymbol{\sigma}} \cdot (\hat{\mathbf{z}} \times \hat{\mathbf{p}}) \delta(z - z_+), \tag{S9}$$

where $\hat{\mathbf{p}}$ is the momentum operator, $\hat{\boldsymbol{\sigma}} = (\hat{\sigma}_x, \hat{\sigma}_y, \hat{\sigma}_z)$ are the Pauli matrices, $\eta$ is the strength of the Rashba SOI, and $\hat{\mathbf{z}}$ is the normal direction of the interface. Eq. (S9) gives rise to a spin-dependent anomalous velocity at the interface: (*21, 52*)

$$v_\alpha'^\beta = -\frac{i}{\hbar}[\mathbf{r}, H_R]_\beta \hat{\sigma}_\alpha = \eta \delta(z - z_+)(\hat{\boldsymbol{\sigma}} \times \hat{\mathbf{z}})_\beta \hat{\sigma}_\alpha \approx \eta \delta(z - z_+)(\mathbf{m} \times \hat{\mathbf{z}})_\beta m_\alpha, \tag{S10}$$

where operator $\hat{\boldsymbol{\sigma}}$ is approximated by the magnetization direction $\mathbf{m}$ at the top surface in contact with FI. $v_\alpha'^\beta$ is the anomalous velocity in $\beta$ direction for spin-$\alpha$. Therefore, the velocity used in the Boltzmann equation Eq. (S1) is modified with replacement $\mathbf{v}_0 \delta_{\alpha,0} \to \mathbf{v}_0 \delta_{\alpha,0} + \mathbf{v}'_\gamma \delta_{\alpha,\gamma}$ at the top surface at $z = z_+$. The anomalous velocity also contributes to the evaluation of charge/spin currents in Eq. (S6). Such an anomalous velocity would not change the drift term in the Boltzmann equation (first term in Eq. (S1)) in the bilayer system studied in this paper. The reason is that the anomalous velocity in Eq. (S10) is in-plane and perpendicular to $\hat{\mathbf{z}}$, while the spatial dependence of $f_\alpha$ is in the $\hat{\mathbf{z}}$ direction, therefore the dot product in the drift term with vanishes identically.

### 6.4 Boltzmann simulation results

With the numerical Boltzmann method described above, we are ready to calculate the longitudinal conductivity and the transverse Hall conductivity as function of magnetization orientation, NM film thickness, and temperature.

We adopt the conventional magnetic field scanning scheme as show in Fig. 3A of main text. We calculate both the longitudinal and transverse resistivity as function of the magnetization angel in the $\alpha$-, $\beta$-, $\gamma$-scans as in Fig. 3A of main text. Then MR is calculated as

$$\mathrm{MR}(\theta) = \frac{\rho(\theta) - \rho_{90°}}{\bar{\rho}} \text{ with } \theta = \alpha, \beta, \gamma. \tag{S11}$$

where $\bar{\rho}$ is the average value of $\rho(\theta)$, and $\rho_{90°}$ is the resistivity value at 90°. Fig. 3A of main text shows the angular dependent of $\rho$ and MR for a Cu film of thickness $t$ = 3 nm, where $\delta$ is chosen to match the average resistivity. It is seen that all three angular dependences are in agreement with the experimental data. The small oscillation in the experimental data for the $\gamma$-scan might be caused by the weak anisotropic MR, which is not included in our simulation. It should note that these angular dependence of $\rho$ and MR ratio in SOMR are identical to that in the SMR. Therefore, it is impossible to tell SOMR from SMR from this angular dependence. The more interesting part is the NM film thickness dependence. As we know for sure that $\rho$ must decrease monotonically as increasing $t$ because of the reducing surface scattering. This is exactly what has been observed experimentally and calculated using the Boltzmann method, as shown in Fig. 4C of main text. In the Boltzmann simulation, we have chosen $\tau_0$ as the bulk Cu relaxation time at room temperature. And $\delta$ is a varying fitting parameter to account for the surface scattering. Using $\delta$ as the only fitting parameter ($\delta$ = 6), we find that the longitudinal resistivity can be fitted with a satisfactory level (Fig. 4C), considering the large error bar and extremely thin film. We also note that the Rashba coupling strength has little effect on the longitudinal resistivity as expected.

It is well-known that the magnitude of SMR depends on the NM film thickness in a non-monotonic fashion, *i.e.*, there is a peak when the film thickness is comparable to the spin diffusion length of NM. However, the SOMR observed in this work shows a monotonic decreasing behavior as increasing NM thickness. We should note that Cu has very long spin diffusion length, much longer than the film thickness. Such monotonic MR ratio can also be fitted using the Boltzmann simulation with only one fitting parameter, *i.e.*, the Rashba coupling constant $\eta$ ($\eta$ = 0.174). Similar to the SMR effect, the SOMR effect also depends quadratically on the spin-orbit coupling strength, therefore $MR \propto \eta^2$.

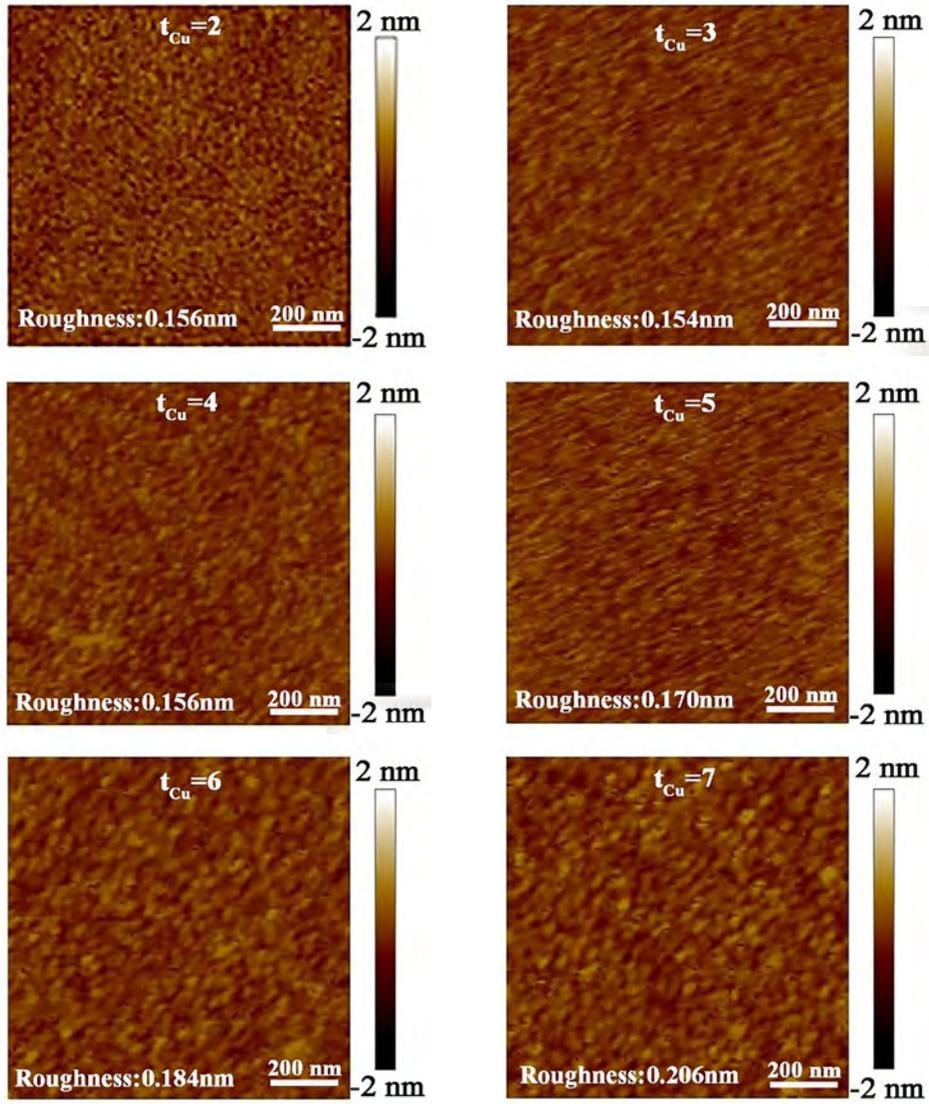

**fig. S1. AFM images of Cu($t_{Cu}$)[Pt(0.4)]/YIG(10)/GGG (111).**

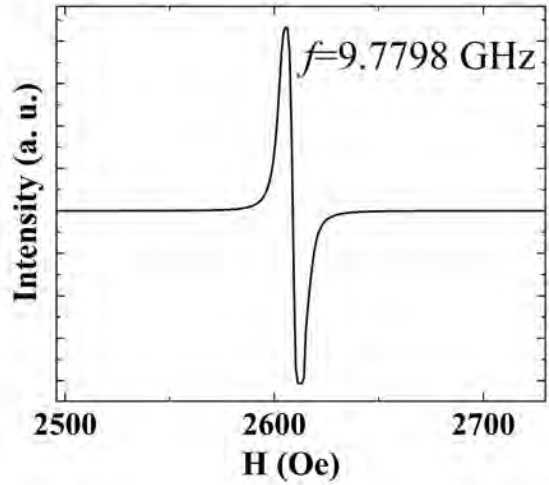

**fig. S2. FMR of the YIG films**. FMR absorption derivative spectrum of YIG(10 nm)/GGG(111) film with field in sample plane measured at room temperature.

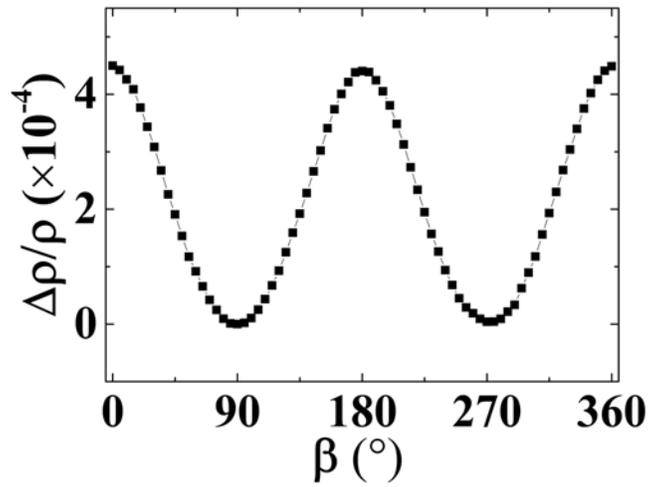

**fig. S3. Spin Hall magnetoresistance in Pt/YIG.** The angular dependent magnetoresistance of our Pt(3.4)/YIG sample measured in $\beta$-scan at room temperature with the magnetic field of 1.5 T.

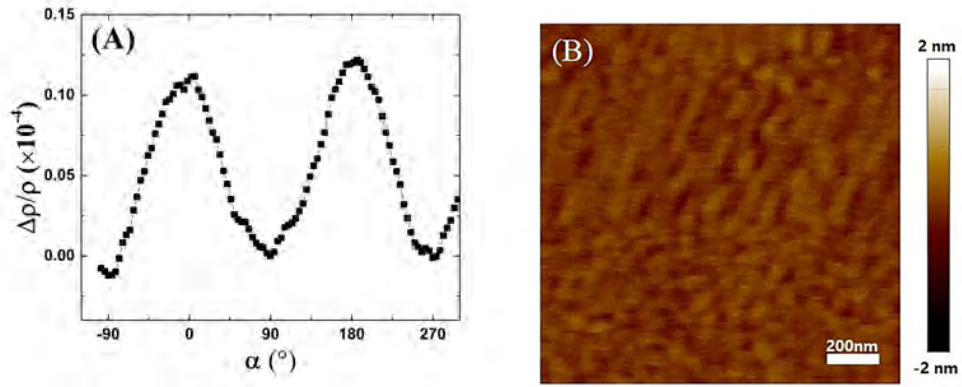

**fig. S4. SMR and AFM image of Au/YIG**. **(A)** The angular dependent MR of the Au(6)/YIG sample measured in $\alpha$-scan with $H$ = 2000 Oe at room temperature. **(B)** AFM image of Au(0.4)/YIG. The rms roughness is 0.135 nm.

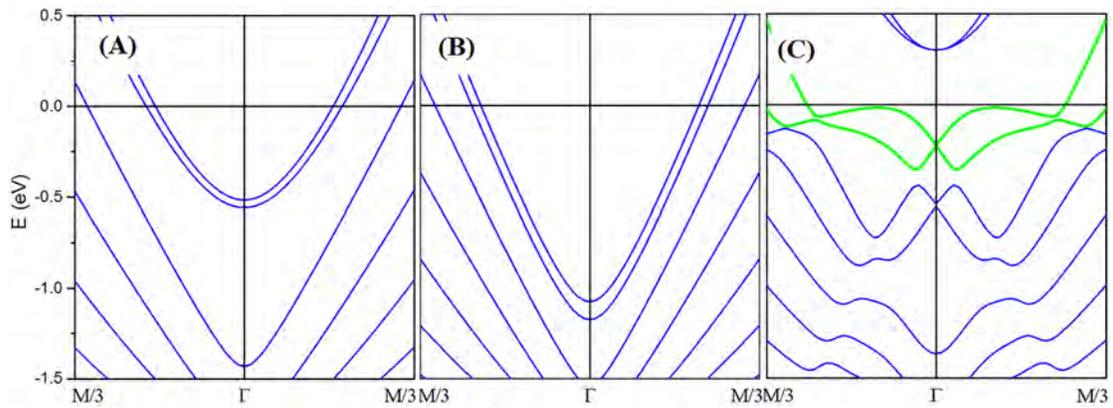

**fig. S5. The band structures of Cu, Au/Cu/Au and Pt/Cu/Pt.** The band structures of **(A)** the Cu ultra-thin film of 14 monolayers, **(B)** the same film covered by Au, **(C)** the same film covered by Pt. The bands marked by green bold lines indicate a Rashba splitting.

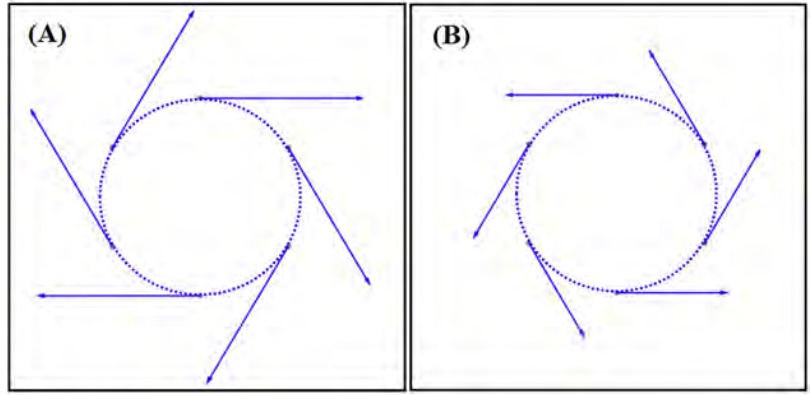

**fig. S6. The spin textures of outer band and inner band.** Spin texture of (**A**) the outer band around -0.35 eV iso-energy surface and (**B**) the inner band around -0.10 eV iso-energy surface. The outer band and inner band are highlighted by green bold lines in Fig. S5C.

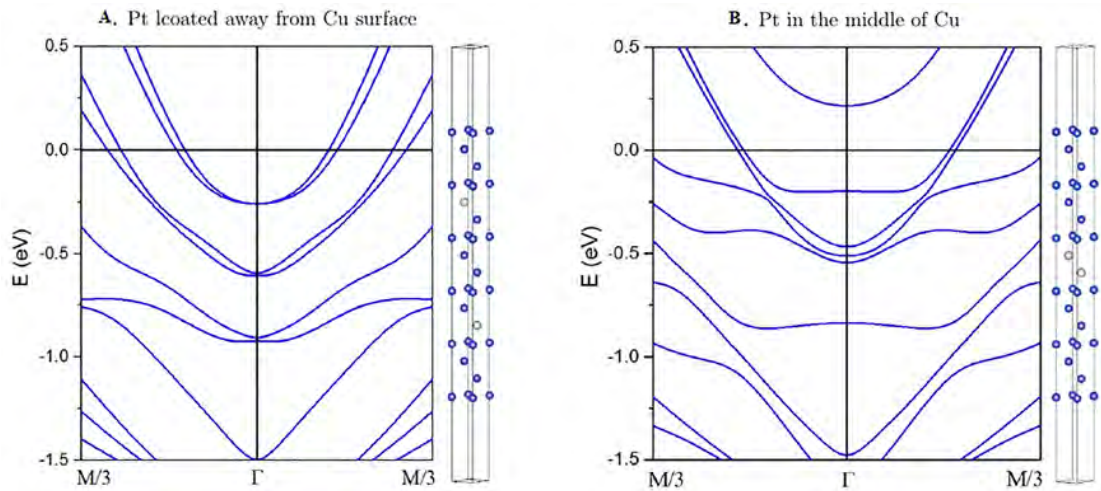

**fig. S7. The Rashba splitting in Cu/Pt/Cu.** The band structures of (**A**) the Cu ultra-thin film of 14 monolayers with Pt located 4 monolayers away from the surface, and it shows a weak Rashba splitting; (**B**) the same film with Pt located in the middle, and there is no Rashba splitting.

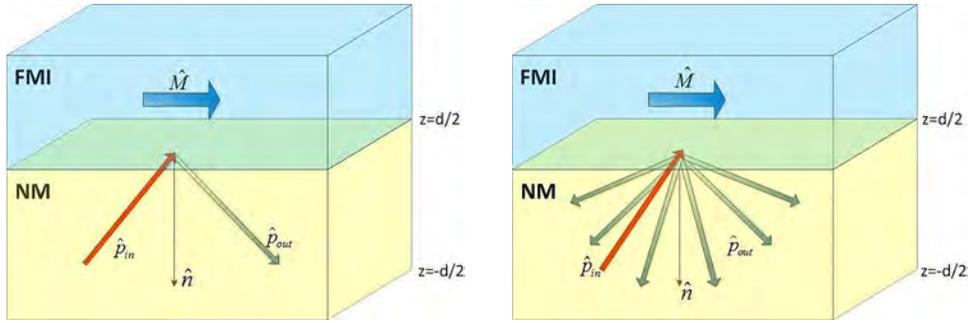

**fig. S8. Specular and diffusive interface scattering in the NM/FI bilayer.**

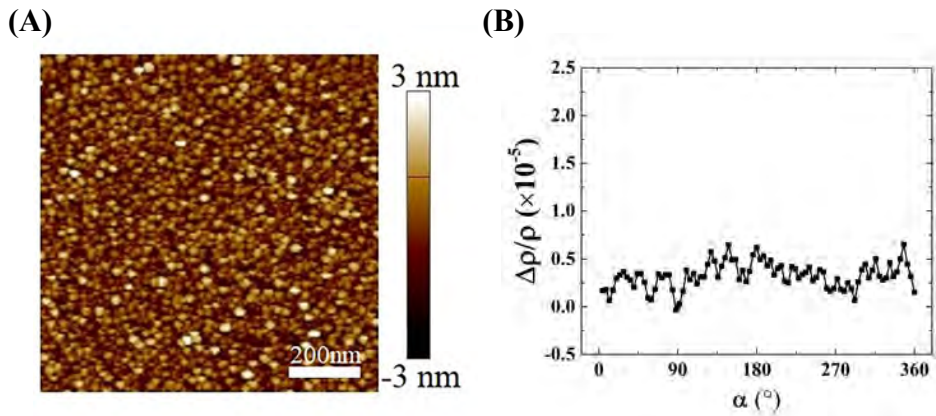

**fig. S9. AFM image of Ag(0.7)/YIG and MR of Cu(3)[Ag(0.7)]/YIG. (A)** AFM image of Ag(0.7)/YIG. The rms roughness is 0.797 nm. **(B)** The angular dependent MR of the Cu(3)[Ag(0.7)]/YIG sample measured in $\alpha$-scan with $H$ = 2000 Oe at room temperature.